\documentclass[american,amssymb,amsmath]{revtex4}
\usepackage[T1]{fontenc}
\usepackage[latin9]{inputenc}
\usepackage{amsmath}
\usepackage{amssymb}
\usepackage{esint}
\usepackage[all, knot]{}
\makeatletter
\@ifundefined{textcolor}{}
{%
 \definecolor{BLACK}{gray}{0}
 \definecolor{WHITE}{gray}{1}
 \definecolor{RED}{rgb}{1,0,0}
 \definecolor{GREEN}{rgb}{0,1,0}
 \definecolor{BLUE}{rgb}{0,0,1}
 \definecolor{CYAN}{cmyk}{1,0,0,0}
 \definecolor{MAGENTA}{cmyk}{0,1,0,0}
 \definecolor{YELLOW}{cmyk}{0,0,1,0}
 }

\makeatother

\usepackage{amstext}
\usepackage{amssymb}
\usepackage[all, knot]{xy}

\begin{document}

\title{Exotic smooth
$\mathbb{R}^4$ and quantum matter}

\author{Jerzy Kr\'ol}

\email{ iriking@wp.pl}

\affiliation{University of Silesia, Institute of Physics, ul. Uniwesytecka 4,
40-007 Katowice}
\begin{abstract}
We follow the point of view that superstring theory, as the theory of quantum gravity in the number of spacetime dimensions bigger than 4, serves as mathematics for both, 4 dimensional QG and exotic smoothness on open 4-manifolds. Extra-dimensions, supersymmetry or some other string techniques, belong to the mathematical toolkit suitable for the above purposes. Physics in dimension 4 is reached via exotic 4-geometries on $\mathbb{R}^4$. In the paper we discuss the techniques of exact superstring backgrounds, CFT and $SU(2)_k$ WZW models, as suitable for the description of effects assigned to the magnetic field and its gravitational backreactions on exotic Euklidean $\mathbb{R}^4$ which is the underlying smoothness for the 4-dimensional spacetime.    
\end{abstract}


\maketitle

\section{Introduction}
A theory of quantum gravity (QG) in 4 dimensions (4d) has not been yet successfully formulated. Any such theory, despite its predictive and calculational power, should serve as explaining the fundamental relation between gravity, (pseudo-)Riemannian geometry and matter on deep quantum level in 4d. How it could be one can imagine based on superstring theory which is the theory of quantum gravity unifying other interactions.  However the 10 spacetime dimensions is unavoidable due to the consistency requirements. Superstring theory represents extremely rich mathematics which surprisingly can be seen as a weakness of the theory. When trying to get 4d physics by compactification of extra dimensions, or by other techniques, one faces huge ambiguity in the choice of the correct background. So maybe superstring theory should be considered as ,,merely'' mathematics but one created especially for the unification and QG purposes. This point of view requires some additional mathematical guidelines but it aims towards physics in 4d. Due to the richness of the mathematics involved in superstring theory it was proposed at ICM2010 \cite{AssKrol2010ICM} that the relation to 4d physical dimensions should go through the mathematical phenomenon of exotic 4-smoothness on open manifolds. The relevant and related problem appears which is considering the standard model of particles and fields (SM) as formulated on 4d Minkowski spacetime but the smoothness of it does not match the 4d smoothness of the theory of gravity. This difference in smooth structures has far reaching consequences. 

The mathematical motivation behind such thinking is that there exists different than standard smoothing on the topological $\mathbb{R}^4$. This takes $\mathbb{R}^4$ to smooth open 4-manifold homeomorphic but non-diffeomorphic to the standard smooth $\mathbb{R}^4$. The standard $\mathbb{R}_{std}^{4}$ is the only differential
structure inherited from the topological product of axes $\mathbb{R}\times\mathbb{R}\times\mathbb{R}\times\mathbb{R}$.
Any non-diffeomorphic smooth $\mathbb{R}^{4}$ is called \emph{exotic} $\mathbb{R}^{4}$. Exotic $\mathbb{R}^{n}$'s exist only for $n=4$. In fact, there exist \emph{infinite continuum many} different exotic $\mathbb{R}^{4}$'s. Here we deal with \emph{small} exotic $\mathbb{R}^{4}$'s which emerge as a result of the failure of the $h$-cobordism theorem in dimension 5 \cite{Asselmeyer2007,Scorpan2005}.
Even though exotic $\mathbb{R}^{4}$'s are smooth 4-manifolds, a big mathematical problem, which constrains also applications to physics, is to find a suitable effective coordinate presentation such that one can do calculus respecting the exoticness of these manifolds (see, however, \cite{Bra:94b,Bra:94a,AsselBrans2011,AsselmeyerKrol2009,AsselmeyerKrol2009a,AssKrol2010ICM,
AsselmKrol2011c,AsselmeyerKrol2011b,AsselmKrol2011d,Ass:96,Krol:04a,Krol:04b,
Krol:2005,Krol2010b,Krol2010b,Krol2010,Sladkowski2001}).    

\section{4d effects from string backgrounds}\label{4d-strings}
Superstring theory (ST) determines its 10d backgrounds. In the case of heterotic ST these are solutions of the following equations of motion at the semiclassical limit of the theory \cite{KK95,AsselmKrol2011f}:
\begin{equation}
\begin{array}{c}
\frac{3}{2}\left[4(\nabla\Phi)^{2}-\frac{10}{3}\square\Phi-\frac{2}{3}R+\frac{1}{12g^{2}}F_{\mu\nu}^{a}F^{a,\mu\nu}\right]=0\\[4pt]
R_{\mu\nu}-\frac{1}{4}H_{\mu\nu}^{2}-\frac{1}{2g^{2}}F_{\mu\rho}^{a}F_{\nu}^{a\rho}+2\nabla_{\mu}\nabla_{\nu}\Phi=0\\[4pt]
\nabla^{\mu}\left[e^{-2\Phi}H_{\mu\nu\rho}\right]=0\\[4pt]
\nabla^{\nu}\left[e^{-2\Phi}F_{\mu\nu}^{a}\right]-\frac{1}{2}F^{a,\nu\rho}H_{\mu\nu\rho}e^{-2\Phi}=0\;.\end{array}\label{EOM}
\end{equation}
with the background fields, metric $G_{\mu\nu}$ leading to the Ricci tensor $R_{\mu\nu}$, strength of the gauge field $F_{\mu\nu}$, antisymmetric 3-form $H_{\mu\nu}$ as strength of the $B$-field and the dilaton $\Phi$.
The fields $F_{\mu\nu}^{a},H_{\mu\nu\rho}$ are usually defined by $F_{\mu\nu}^{a}=\partial_{\mu}A_{\nu}-\partial_{\nu}A_{\mu}+f^{abc}A_{\mu}^{b}A_{\nu}^{c}$
and $H_{\mu\nu\rho}=\partial_{\mu}B_{\nu\rho}-\frac{1}{2g^{2}}\left[A_{\mu}^{a}F_{\nu\rho}^{a}-\frac{1}{3}f^{abc}A_{\mu}^{a}A_{\nu}^{b}A_{\rho}^{c}\right]+{\rm permutations}$.
The constants $f^{abc}$ are structure constants of the gauge group
and $A_{\mu}^{a}$ is the effective gauge field. Because of that one can derive EOM (\ref{EOM}) from the effective 4d action: 
\begin{equation}
\begin{array}{c}

S=\int d^{4}x\sqrt{G}e^{-2\Phi}[R+4(\nabla\Phi)^{2}-\frac{1}{12}H^{2} 
-\frac{1}{4g^{2}}F_{\mu\nu}^{a}F^{a,\mu\nu}+\frac{C}{3}]
\end{array}
\end{equation}
we set $g_{str}=1$ and for the gauge coupling $g^{2}=2/k_{g}$. $C$ is the l.h.s. of the first equation in (\ref{EOM}). 
On such geometric 10d backgrounds one defines the superconformal 2d quantum field theory (CFT) and generates correlation functions etc. of string theory by the inclusion of the corresponding vertex operators \cite{Segal2001}. One way to define CFT on a background is to consider $\sigma$-model with this background as a target. 
The CFT in question is the worldsheet superconformal ${\cal N}=4$ $\hat{c}=4$ one. The suplementary approach relies on finding the representations of this algebra on the background and trying to relate it with the $\sigma$-model. The superconformal coordinates are defined in terms of the currents and fields of the $\sigma$-model. The evaluation of the spectra of the theory is from the one side, in terms of modular invariant characters of the group manifold, and in terms of currents of the $\sigma$-model, on the other. These are considered as complementary to each other. When the background is factored into say $S^3\times \mathbb{R}\times M^6$ the correlators of the CFT should respect this with eventual reduction of the supersymmetries to ${\cal N}=2$. 

To be more specific let us, following \cite{Antoniadis94}, consider the superconformal algebra ${\cal N}=4$ $\hat{c}=4$ which is defined by the stress energy tensor $T(z)$, the supercurrents $G_a(z), a=1,2,3,4$ and $SU(2)_k$ Kac-Moody currents at the level $k$, i.e. $S_i(z), i=1,2,3$. The following OPE relations emerge by closing the algebra:
 \begin{equation}
\begin{array}{c}
T(z)T(w)\sim \frac{3\hat{c}}{4(z-w)^4}+\frac{2T(w)}{(z-w)^2}+\frac{\partial T(w)}{(z-w)}\\[5pt]
T(z)G_a(w)\sim \frac{3G_a(w)}{2(z-w)^2}+\frac{\partial G_a(w)}{(z-w)} \\[5pt]
T(z)S_i(w)\sim \frac{S_i(w)}{2(z-w)^2}+\frac{\partial S_i(w)}{(z-w)} \\ [5pt]
G_i(z)G_j(w)\sim \delta_{ij}\frac{\hat{c}}{(z-w)^3}-4\epsilon_{ijl}\frac{S_l(w)}{(z-w)^2}+\delta_{ij}\frac{2T(w)}{(z-w)}, i,j,l=1,2,3,4 \\[5pt]
S_i(z)G_j(w)\sim \frac{1}{2(z-w)}(\delta_{ij}G_4(w)+\epsilon_{ijl}G_l(w)),i,j,l=1,2,3,4\\[5pt]
S_i(z)S_j(w)\sim -\delta_{ij}\frac{n}{2(z-w)^2}+\epsilon_{ijl}\frac{S_l(w)}{(z-w)}
\end{array}\label{SU-Con-Alg}
\end{equation}
where $n=1$ for $\hat{c}=4$. 
Next we turn to the exact realization of the above algebra in terms of the $SU(2)_k\times U(1)_Q$ bosonic currents $J_a,a=1,2,3$, and their superpartners which are free fermionic fields $\Psi^a$. The result reads:
\begin{equation}
\begin{array}{c}
T=-\frac{1}{2}[\frac{2}{k+2}J_i^2+J^2_4-\Psi_a \partial \psi_a+Q\partial J_4]\\[5pt]
G_4= \sqrt{\frac{2}{k+2}}(J_i\Psi_i+\frac{1}{3}\epsilon_{ijl}\Psi_i\Psi_j\Psi_l) +J_4\Psi_4+Q\partial \Psi_4 \\[5pt]
G_i= \sqrt{\frac{2}{k+2}}(J_i\Psi_4+\epsilon_{ijl}J_j\Psi_l+\epsilon_{ijl}\Psi_4\Psi_j\Psi_l) +J_4\Psi_i+Q\partial \Psi_i,i=1,2,3,4, \\[5pt]
S_i=\frac{1}{2}(\Psi_4\Psi_i+\frac{1}{2}\epsilon_{ijl}\Psi_j\Psi_l)\;.
\end{array}\label{SU-Con-Alg}
\end{equation}

Next we complexify the generators and bosonise the free fermions by the scalar fields $H^+,H^-$:
\begin{equation}
\begin{array}{c}
T=-\frac{1}{2}[(\partial H^+)^2+(\partial H^-)^2+Q^2(J_1^2+J_2^2+J_3^2) +J_4^2+Q\partial J_4] \\[5pt]
G=\frac{G_1+iG_2}{\sqrt{2}}=-(\Pi_k^{\dag}e^{-\frac{i}{\sqrt{2}}H^-}+P_k^{\dag}e^{\frac{i}{\sqrt{2}}H^-})e^{\frac{i}{\sqrt{2}}H^+} \\[5pt]
\tilde{G}=\frac{G_4+iG_3}{\sqrt{2}}=(\Pi_k^+e^{\frac{i}{\sqrt{2}}H^-}-P_ke^{-\frac{i}{\sqrt{2}}H^-})e^{\frac{i}{\sqrt{2}}H^+} \\[5pt]
S_3=\frac{1}{\sqrt{2}}\partial H^+\:,\;S_{\pm}=e^{\pm i\sqrt{2}H^+}
\end{array}\label{SU-Con-Alg}
\end{equation}
where the coordinate currents are as follows:
\begin{equation}
\begin{array}{c}
P_k=Q(J_1+iJ_2) \\[5pt]
P_k^{\dag}=Q(-J_1+iJ_2) \\[5pt]
\Pi_k=J_4+iQ(J_3+\sqrt{2}\partial H^-) \\[5pt]
\Pi_k^{\dag}=-J_4+iQ(J_3+\sqrt{2}\partial H^-)\;.
\end{array}\label{SU-Con-Alg}
\end{equation}
This gives the correct change of coordinates and realization of the ${\cal N}=4,\hat{c}=4$ super CFT in terms of $\sigma$-models currents and fields on $SU(2)_k\times \mathbb{R}_{\phi}\times W^6$.
This background realizes 4d part $SU(2)_k\times \mathbb{R}_{\phi}$ as curved 4-manifold. It appears naturally when starting with flat background $\mathbb{R}^4\times W^6$ and almost constant magnetic field is switched on on $\mathbb{R}^4$ part. In closed string theory even constant magnetic field causes the background to be curved and the correct choice is as above \cite{KK95}. On the level of representation of ${\cal N}=4,\hat{c}=4$ super CFT algebra given by the $\sigma$-model with target $\mathbb{R}^4\times W^6$ one has, non-modified by background curved fields, operators:
\begin{equation}
\begin{array}{c}
T=-\frac{1}{2}[(\partial H^+)^2+(\partial H^-)^2-PP^{\dag}-\Pi\Pi^{\dag}] \\[5pt]
G=\frac{G_1+iG_2}{\sqrt{2}}=-(\Pi^{\dag}e^{-\frac{i}{\sqrt{2}}H^-}+P^{\dag}e^{\frac{i}{\sqrt{2}}H^-})e^{\frac{i}{\sqrt{2}}H^+} \\[5pt]
\tilde{G}=\frac{G_4+iG_3}{\sqrt{2}}=(\Pi^+e^{\frac{i}{\sqrt{2}}H^-}-Pe^{-\frac{i}{\sqrt{2}}H^-})e^{\frac{i}{\sqrt{2}}H^+} \\[5pt]
S_3=\frac{1}{\sqrt{2}}\partial H^+\:,\;S_{\pm}=e^{\pm i\sqrt{2}H^+}\;.
\end{array}\label{Flat-1}
\end{equation}
Here, \begin{equation*}
\begin{array}{c}
P=J_1+iJ_2,\;\; P^{\dag}=-J_1+iJ_2 \\[5pt]
\Pi=J_4+iJ_3\;\; \Pi^{\dag}=-J_4+iJ_3
\end{array}\label{Flat-2}
\end{equation*}
and $J_a=\partial \Phi_a,\,a=1,2,3,4$ are bosonic $U(1)$-currents, and free fermions are again written in terms of two bosons, $H^+,H^-$. Also, the decomposition of the $SO(4)_1$ fermionic currents, $\Psi_i\Psi_j$ in terms of two $SU(2)_1$ currents $S_i,S_k$ was performed, which reads:
\begin{equation*}
\begin{array}{c}
S_i=\frac{1}{2}(\Psi_4\Psi_i+\frac{1}{2}\epsilon_{ijk}\Psi_j\Psi_l)\to (\frac{1}{2}\partial H^+,e^{\pm i\sqrt{2}H^+}) \\[5pt]
S_i=\frac{1}{2}(-\Psi_4\Psi_i+\frac{1}{2}\epsilon_{ijk}\Psi_j\Psi_l)\to (\frac{1}{2}\partial H^-,e^{\pm i\sqrt{2}H^-})\;.
\end{array}\label{Flat-3}
\end{equation*}
Finally, one arrives at (\ref{Flat-1}) for flat $\mathbb{R}^4$ in the background $\mathbb{R}^4\times W^6$. Thus, the inclusion of the magnetic field on $\mathbb{R}^4$ results in the shift in the heterotic string background as follows:
\begin{equation}\label{bckgrs}
\mathbb{R}^4\times W^6\longrightarrow SU(2)_k\times \mathbb{R}_{\phi}\times W^6\,. 
\end{equation}
The 4d geometry is thus shifted in accord with:
\begin{equation}\label{4-bc}
\mathbb{R}^4\longrightarrow SU(2)_k\times \mathbb{R}_{\phi}\,. 
\end{equation} This is certainly achieved under the presence of 10d supersymmetry, however the supersymmetry we consider as necessary technical condition allowing for the mathematical description of the shift as above. The shift where also topology of the background is changed, is protected by supersymmetry. We do not, however, assign any real existence to such understood supersymmetry. Rather we are interested in the 4d geometry which would correspond to the shifted one, as in (\ref{4-bc}), and which would have physical meaning as underlying geometry for 4-spacetime. 

Given this explicit realization of ${\cal N}=4$ algebra as CFT, one can construct modular invariant combinations respecting the ${\cal N}=4$ superconformal symmetry. The 10d spacetime target ${\cal N}=4$ supersymmetry is thus induced which guarantees the stability of the solutions against string $\alpha$' corrections. From the point of view of 4 noncompact dimensions there exist 2 covariantly constant spinor fields in the heterotic string background.  These are BRST-invariant ${\cal N}=4$ spins:
\begin{equation}\label{spin-1}  
 \Theta_{\pm}=e^{\frac{i}{\sqrt{2}}(H_1^+{\pm}H^+_2)}\;.
\end{equation}
The level-1 character combinations associated with the $SU(2)_{H^+}$'s, after GSO projection, read:
\begin{equation}  
 \frac{1}{2}(1-(-1)^{l_1+l_2})\chi^{l_1}_{H_1^+}\chi^{l_2}_{H_2^+}=\chi^{l_1}_{H_1^+}\chi^{1-l_1}_{H_2^+}\delta_{l_2,1-l_1}
\end{equation}
where $l_A=1$ corresponds to spin-$\frac{1}{2}$ character of the $SU(2)_1$ Kac-Moody algebra, and $l_A=0$ to spin-0 character of this. 
These characters have to be combined with $W^6$ characters and $U(1)_Q$ ones ($Q$ is the charge of the linear dilaton from $\mathbb{R}_{\phi}$ direction of the background). We are choosing flat Minkowski $M^6$ as the $W^6$ part. 
On the level of partition function we have flat toruses directions: $T^{(2)}\times T^{(4)}$. The $\mathbb{Z}_2$ orbifolding of the background $\mathbb{R}^4\times T^{(2)}\times T^{(4)}$ is extended over $SU(2)_k\times \mathbb{R}_{\phi}\times T^{(2)}\times T^{(4)}$ such that it non-trivially mixes the $\Theta$ and $\tilde{\Theta}$-spins leaving the physical ones as in (\ref{spin-1}). The orbifolded partition function in the type II case, reads \cite{KK95}:
\begin{equation}\label{Z1}
Z^{\mathbb{Z}_2}_{W}={\rm Im}\tau^{\frac{1}{2}}|\eta|^2\frac{1}{2V}\sum_{\gamma,\delta}Z_{so(3)}[^{\gamma}_{\delta}]\frac{1}{8}\sum_{ ^{\alpha,\beta,\overline{\alpha}}_{\overline{\beta},h,g}}(-1)^{(\alpha+\overline{\alpha})(1+\delta)+\beta+\overline{\beta}}\frac{\vartheta^2\left[^{\alpha}_{\beta}\right]}{\eta^2}\frac{\vartheta^2\left[^{\alpha+h}_{\beta+g}\right]}{\eta^2}\frac{\overline{\vartheta}^2\left[^{\overline{\alpha}}_{\overline{\beta}}\right]}{\overline{\eta}^2}\frac{\vartheta^2\left[^{\overline{\alpha}+h}_{\overline{\beta}+g}\right]}{\overline{\eta}^2} Z_2\left[^{0}_0\right]Z_4\left[^{h}_g\right]\,.
\end{equation}
Here $Z_2\left[^{0}_0\right]=\frac{\Gamma (2,2)}{|\eta|^4}$ from the $T^{(2)}$ compactification, and $Z_4\left[^{h}_g\right]=\frac{|\eta|^4}{|\vartheta[^{1+h}_{1+g}]\vartheta[^{1-h}_{1-g}]|^2}$ from the $T^{(4)}$ one, $V=\frac{(k+2)^{3/2}}{8\pi}$ is the volume of $S^3$ at the level $k$. The level $k$ $\vartheta$-functions are defined as usual by: 
\begin{equation}
\vartheta_{m,k}(\tau,\nu)=\sum_{n\in \mathbb{Z}}{\rm exp}\left[2\pi ik\left(n+\frac{m}{2k}\right)^2\tau - 2\pi ik\left(n+\frac{m}{2k}\right)^2\nu\right]\,,
\end{equation}
and the following combination of characters of $SU(2)_k$ was used:
\begin{equation}
Z_{so(3)}[^{\alpha}_{\beta}]=e^{-i\pi \alpha\beta k/2}\sum^k_{l=0}e^{i\pi \beta l}\chi_l\overline{\chi}_{(1-2\alpha)l+\alpha k}\,.
\end{equation} 
The appropriate heterotic partition function is obtained from (\ref{Z1}) by the substitutions: 
\begin{equation}
(-1)^{\overline{\alpha}+\overline{\beta}}\frac{\overline{\vartheta}^2\left[^{\overline{\alpha}}_{\overline{\beta}}\right]}{\overline{\eta}^2}\longrightarrow \frac{\overline{\vartheta}^6\left[^{\overline{\alpha}}_{\overline{\beta}}\right]}{\overline{\eta}^6}\frac{1}{2}\sum_{\gamma,\delta=0}^1\frac{\overline{\vartheta}^8\left[^{\gamma}_{\delta}\right]}{\overline{\eta}^8}
\end{equation}
for $O(12)\otimes E_8$ case, and 
\begin{equation}
(-1)^{\overline{\alpha}+\overline{\beta}}\frac{\overline{\vartheta}^2\left[^{\overline{\alpha}}_{\overline{\beta}}\right]}{\overline{\eta}^2}\longrightarrow \frac{\overline{\vartheta}^{14}\left[^{\overline{\alpha}}_{\overline{\beta}}\right]}{\overline{\eta}^{14}}
\end{equation}
for the $O(28)$ case.

In that way one obtains the modular invariant partition functions $Z^W(\tau,\overline{\tau})$ for the $SU(2)_k\times \mathbb{R}_{\phi}\times M^6$, $k$ even, heterotic background and can compare with the flat one $Z_0(\tau,\overline{\tau})$ for $\mathbb{R}^4\times M^6$:
\begin{equation}\label{Z-W}
Z^W(\tau,\overline{\tau})={\rm Im}\tau^{3/2}|\eta|^6\frac{\Gamma(SO(3)_{k/2})}{V}Z_0(\tau,\overline{\tau})\,.
\end{equation}  
Here $\Gamma(SO(3)_{k/2})=1/2\sum^1_{\gamma,\delta}Z_{so(3)}[^{\gamma}_{\delta}]$ is the partition function of the $SO(3)$ WZW model at the level $k/2$. The factorization as above allows for grasping the effects assigned particularly to the change of the 4d part, according to (\ref{4-bc}). 

\section{2d CFT from the end of small Exotic $\mathbb{R}^4$}\label{2dCFT}
As we already mentioned the standard $\mathbb{R}_{std}^{4}$ is the only smooth differential structure which agrees with the topological product of axes $\mathbb{R}\times\mathbb{R}\times\mathbb{R}\times\mathbb{R}$.
An exotic $\mathbb{R}^{4}$ is the same topological 4-manifold  $\mathbb{R}^{4}-$ but with a different (i.e. non-diffeomorphic) smooth structure. This is possible only for $\mathbb{R}^{4}$ which is the only Euclidean space $\mathbb{R}^{n}$ with an exotic smoothness structure \cite{Asselmeyer2007,Scorpan2005}.
  
One can relate these 4-exotics with some structures on $S^{3}$ (see e.g. \cite{AsselmeyerKrol2009,AsselmeyerKrol2011,AsselmKrol2011c}) provided it is placed at the boundary of a compact contractible 4-submanifold -- the Akbulut cork. If so, one can prove that exotic smoothness of the $\mathbb{R}^{4}$ is tightly related with codimension-one foliations of this $S^{3}$, hence with the 3-rd real cohomology classes of $S^{3}$. In this sense we classify exotic smooth $\mathbb{R}^{4}$'s, from the so called \emph{radial family}, by $H^3(S^3,\mathbb{R})$ \cite{AsselmeyerKrol2009,AsselmeyerKrol2011b}. 

Small exotic $\mathbb{R}^{4}$ is determined by the compact 4-manifold $A$ with boundary $\partial A$ which is homology 3-sphere, and attached several \emph{Casson handles} CH's. $A$ is the Akbulut cork and CH is built from many stages towers of immersed 2-disks. These 2-disks cannot be embedded and the intersection points can be placed in general position in 4D in separated double points. Every CH has infinite many stages of intersecting disks. However, CH is topologically the same as (homeomorphic to) open 2-handle, i.e. $D^2\times \mathbb{R}^2$. Now if one replaces CH's, from the above description of small exotic $\mathbb{R}^{4}$, by ordinary open 2-handles (with suitable linking numbers in the attaching regions) the resulting object is standard $\mathbb{R}^{4}$. The reason is the existence of infinite (continuum) many diffeomorphism classes of CH, even though all are topologically the same.

In the case of integral \emph{$H^{3}(S^{3},\mathbb{Z})$}
one yields the relation of exotic $\mathbb{R}_{k}^{4}$, $k[\:]\in H^{3}(S^{3},\mathbb{Z})$, $k\in\mathbb{Z}$ with the WZ term of the $k$ WZW model on $SU(2)$.
This is because the integer classes in $H^{3}(S^{3},\mathbb{Z})$
are of special character. Topologically, this case refers to flat
$PSL(2,\mathbb{R})-$bundles over the space $(S^{2}\setminus\left\{ \mbox{\mbox{k} punctures}\right\} )\times S^{1}$
and due to the Heegard decomposition one obtains the relation \cite{AsselmeyerKrol2009}:\begin{equation}
\frac{1}{(4\pi)^{2}}\langle GV(\mathcal{F}),[S^{3}]\rangle=\frac{1}{(4\pi)^{2}}\,\intop_{S^{3}}GV(\mathcal{F})=\pm(2-k)\label{eq:integer-GV}\end{equation}
the sign depends on the orientation of the fundamental class $[S^{3}]$.
We can interpret the Godbillon-Vey invariant of the foliation of $S^3$ as WZ term.
Namely we consider a smooth map $G:S^{3}\to SU(2)$ and 3-form
$\Omega_{3}=Tr((G^{-1}dG)^{3})$ so that the integral\[
\frac{1}{8\pi^{2}}\intop_{S^{3}=SU(2)}\Omega_{3}=\frac{1}{8\pi^{2}}\intop_{S^{3}}Tr((G^{-1}dG)^{3})\in\mathbb{Z}\]
is the winding number of $G$. Thus indeed every Godbillon-Vey class with integer
value like (\ref{eq:integer-GV}) is generated by a 3-form $\Omega_{3}$.
Therefore the Godbillon-Vey class is the WZ term of the $SU(2)_{k}$ WZW model. The foliation of $S^3$ with this GV class is generated by some exotic $\mathbb{R}^4$, namely $\mathbb{R}^4$. Thus, we see that \emph{the structure of exotic $\mathbb{R}_{k}^{4}$'s, $k\in\mathbb{Z}$ from the radial family determines the WZ term of the $k-2$ WZW model on $SU(2)$.}

This WZ term is required by the cancellation of the quantum anomaly
due to the conformal invariance of the classical $\sigma$-model on
$SU(2)$. Thus we have a way how to obtain this cancellation term
from smooth 4-geometry: when a smoothness of the ambient 4-space,
in which $S^{3}$ is placed as a part of the boundary of the cork,
is precisely the smoothness of exotic $\mathbb{R}_{k}^{4}$, then the WZ term of the classical $\sigma$-model with target $S^{3}=SU(2)$, i.e. $SU(2)_{k}$ WZW,
is generated by this 4-smoothness. The important correlation follows:

\emph{The change of smoothness of exotic $\mathbb{R}_{k}^{4}$ to exotic
$\mathbb{R}_{l}^{4}$, $k,\, l\in\mathbb{Z}$ both from the radial
family, corresponds to the change of the level $k$ of the WZW model
on $SU(2)$, i.e. $k\,{\rm WZW\to}l\,{\rm WZW}$.}

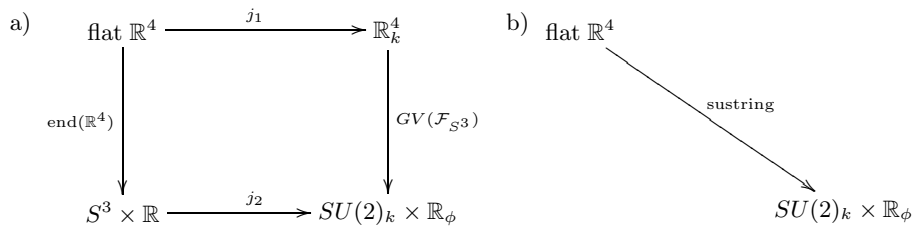
\begin{figure}[ht]
\centering
\begin{tabular}{cc}
a) \xymatrix{
   \rm{flat}\;\mathbb{R}^4
     \ar[rr]^{j_1}
     \ar[dd]_{\rm{end(\mathbb{R}^4)}} 
     && \mathbb{R}^4_k
     \ar[dd]^{GV({\cal{F}}_{S^3})} \\ \\
      S^3\times \mathbb{R}
     \ar[rr]^{j_2} 
      && SU(2)_k\times \mathbb{R}_{\phi} }  & \label{diag2}
b) \xymatrix{
\rm{flat}\; \mathbb{R}^4
\ar[ddrr]^{\rm{sustring}}
\\ \\ & & SU(2)_k\times \mathbb{R}_{\phi}}
\end{tabular}
\caption{a) $j_1$ is the change of the standard smooth $\mathbb{R}^4$ to the exotic $\mathbb{R}^4_k$, $\rm{end(\mathbb{R}^4)}$ assigns the standard end to $\mathbb{R}^4$, $GV({\cal{F}}_{S^3})$ generates the $\rm{WZ}_k$-term from exotic $\mathbb{R}^4_k$ via $GV$ invariant of the codim.-1 foliation of $S^3$. b) The change of string backgrounds s.t. flat $\mathbb{R}^4$ part is replaced by the linear dilaton background $SU(2)_k\times \mathbb{R}_{\phi}$.}\label{fig-1}
\end{figure}

The end of the exotic $\mathbb{R}_{k}^{4}$ i.e.
$S^{3}\times\mathbb{R}$ cannot be standard smooth and it
is in fact fake smooth $S^{3}\times_{\Theta_{k}}\mathbb{R}$, \cite{Fre:79}.
So we have determined, via WZ term, the geometry of $SU(2)_{k-2}\times\mathbb{R}$
as corresponding to the exotic geometry of the end of $\mathbb{R}_{k}^{4}$.
Thus, the change of smoothness on $\mathbb{R}^4$, from standard to exotic, corresponds to the change of the geometry of the end, from $S^3\times \mathbb{R}$ to $SU(2)_k\times \mathbb{R}$. This last, however, emerges as the geometry of the exact string background as we discussed in the last paragraph. We can illustrate this correspondence by the diagram as in Fig. (\ref{fig-1}). 
 
\section{Spectra of particles in exotic 4-geometry in spacetime}
We analyze the situation where smoothness of spacetime, as a 4-manifold, is rather exotic $\mathbb{R}^4$ than standard flat one. Quantum particles, considered as test particles, should show modified interactions, hence energy spectra, when in this exotic structures. One important observation is in order: exotic $\mathbb{R}^4$ cannot be flat; if it were it would have to be standard $\mathbb{R}^4$. If so, in some regime gravity assigned to such curvature should be considered in terms of quantum gravity rather than classical general relativity. However, QG is not a working complete theory in 4d. In fact, it does not exist in 4d. That is why we refer to the relation of exotic $\mathbb{R}^4$ with background of string theory from Secs. \ref{4d-strings} and \ref{2dCFT}, and make use of superstring techniques such that 4d results are derivable.  

Following this philosophy and Refs. \cite{AsselmKrol2011f,KK95}, let us switch on strong, almost constant magnetic field on $\mathbb{R}^4$ and respect its gravitational backreaction as acting upon some spectra of test particles. The gravitational backreaction of the magnetic field is written as the curvature of $SU(2)_k\times \mathbb{R}_{\phi}$ replacing flat $\mathbb{R}^4$ and the presence of supersymmetry is crucial here. At quantum regime further gravitational effects are grasped via the \emph{marginal} deformations of the corresponding superconformal WZW model on $SU(2)_k$. 
Thus, the flat 4d background is curved due to the presence of, say, magnetic field. At the deep quantum regime one recognizes surviving 4d geometry as the curved geometry of the string background $SU(2)_k\times \mathbb{R}_{\phi}$. However, the nontrivial change of coordinates is performed such that 4-non-compact dimensions are now represented by superconformal fields, as in Sec. \ref{4d-strings}. When this is achieved, insertion of magnetic field, at even deeper quantum regime, is not described by further curving of the spacetime manifold. Rather, the deformation of 2d CFT is in order to calculate relevant expressions. This is the strategy we follow in this section.    

Thus we consider two kinds of marginal deformations of the supersymmetric WZW model on $SU(2)_k\times \mathbb{R}_{\phi}$: magnetic and gravitational.  In the case
of a single magnetic field $F$ the operators corresponding to truly marginal deformations and in the case of the current-current interactions, are given by the bilinear product of currents \cite{KK95}:
$V_{F}=F\frac{(J^{3}+\psi^{1}\psi^{2})}{\sqrt{k+2}}\frac{\overline{J}}{\sqrt{k_{g}}}$
where $J^{3}$, $\overline{J}^{3}$ are the $SU(2)$ currents, $J$,
$\overline{J}$ are holomorphic and antiholomorphic ones, and the
right moving current $\overline{J}$ is normalized as $<\overline{J}(1)\overline{J}(0)>=k_{g}/2$.
The corresponding gravitational deformation reads: $V_{gr}={\cal R}\frac{(J^{3}+\psi^{1}\psi^{2})\overline{J}^{3}}{\sqrt{k+2}\sqrt{k}}$.

Let us include these marginal deformations $V_{F}$ and $V_{gr}$
as $O(1,1)$ boost in the lattice of charges of the theory. The effects
will be encoded in the zero-modes of the $SU(2)_{k}$ currents, $J^{3}$,
$\overline{J}^{3}$, i.e. $I$, $\bar{I}$, the zero-modes of the
holo- (antiholo-)morphic currents, $J$, $\overline{J}$, i.e. ${\cal P}$,
$\bar{{\cal P}}$, and the zero-mode of the holomorphic helicity current,
$\psi^{1}\psi^{2}$, which is denoted by ${\cal Q}$. Then the zero-modes
of the algebra are:

\[
L_{0}=\frac{{\cal Q}^{2}}{2}+\frac{I^{2}}{2}+...\,,\;\bar{L}_{0}=\frac{\bar{{\cal P}}^{2}}{k_{g}}+...\]
 which gives rise to the relevant for the $V_{F}$ perturbation part:

\begin{equation}
L_{0}=\frac{({\cal Q}+I)^{2}}{k+2}+\frac{k}{2(k+2)}\left({\cal Q}-\frac{2}{k}I\right)^{2}+...\;.\end{equation}
 The $O(1,1)$ boost mixes the holomorphic zero-mode current $I+{\cal Q}$
with the antiholomorphic $\bar{{\cal P}}$:

\begin{equation}
\begin{array}{c}
L'_{0}=\left(\cosh x\frac{{\cal Q}+I}{\sqrt{k+2}}+\sinh x\frac{\bar{{\cal P}}}{\sqrt{k_{g}}}\right)^{2}+\frac{k}{2(k+2)}\left({\cal Q}-\frac{2}{k}I\right)^{2}+...\\[4pt]
\bar{L}'_{0}=\left(\sinh x\frac{{\cal Q}+I}{\sqrt{k+2}}+\cosh x\frac{\bar{{\cal P}}}{\sqrt{k_{g}}}\right)^{2}+...\;.\end{array}\label{Ap2}\end{equation}
Next we include the $V_{gr}$ deformation. This deformation is symmetric
hence we perform in addition to $V_{F}$ the $O(2)$ transformation
which mixes the antiholomorphic $\bar{I}$, $\bar{{\cal P}}$. After
substituting $F=\sinh(2x)$ we have the following perturbation $\delta L_{0}=L'_{0}-L_{0}$
in $L_{0}$ (and $\bar{L}_{0}$):

\begin{equation}
\delta L_{0}=\left[\frac{{\cal R}\bar{I}}{\sqrt{k}}+\frac{F^{2}}{\sqrt{k_{g}}}\right]\frac{{\cal Q}+I}{\sqrt{k+2}}+\left(\sqrt{1+{\cal R}^{2}+F^{2}}-1\right)\left[\frac{({\cal Q}+I)^{2}}{2(k+2)}+\frac{1}{2({\cal R}^{2}+F^{2})}\left(\frac{{\cal R}\bar{I}}{\sqrt{k}}+\frac{F\bar{{\cal P}}}{\sqrt{k_{g}}}\right)^{2}\right].\end{equation}

This perturbation gives the mass spectra $L_{0}=M_{L}^{2}$ ($\bar{L}_{0}=M_{R}^{2}$):

\begin{equation}
\begin{array}{c}
M_{L}^{2}=-\frac{1}{2}+\frac{{\cal Q}^{2}}{2}+\frac{1}{2}\sum_{i=1}^{3}{\cal Q}_{i}^{2}+\frac{(j+1/2)^{2}-({\cal Q}+I)^{2}}{k+2}+E_{0}+\\[4pt]
+\frac{1+\sqrt{1+F^{2}}}{2}\left[\frac{{\cal Q}+I}{\sqrt{k+2}}+\frac{F\bar{{\cal P}}}{\sqrt{k_{g}}\left(1+\sqrt{1+F^{2}}\right)}\right]^{2}+\\[4pt]
+\frac{1+\sqrt{1+{\cal R}^{2}}}{2}\left[\frac{{\cal Q}+I}{\sqrt{k+2}}+\frac{{\cal R}\bar{I}}{\sqrt{k}(1+\sqrt{1+{\cal R}^{2}})}\right]^{2}\end{array}\label{Ap3}\end{equation}
and with the help of the following gravitational backreaction moduli:

\begin{equation}
\lambda=\sqrt{{\cal R}+\sqrt{1+{\cal R}^{2}}},\frac{1}{\lambda}=\sqrt{-{\cal R}+\sqrt{1+{\cal R}^{2}}}\label{eq:moduli-lambda}\end{equation}
 the spectra can be obtained:

\begin{equation}
\begin{array}{c}
M_{L}^{2}=-\frac{1}{2}+\frac{{\cal Q}^{2}}{2}+\frac{1}{2}\sum_{i=1}^{3}{\cal Q}_{i}^{2}+\frac{(j+1/2)^{2}-({\cal Q}+I)^{2}}{k+2}+E_{0}+\\[4pt]
+\frac{1+\sqrt{1+F^{2}}}{2}\left[\frac{{\cal Q}+I}{\sqrt{k+2}}+\frac{F\bar{{\cal P}}}{\sqrt{k_{g}}\left(1+\sqrt{1+F^{2}}\right)}\right]^{2}+\\[4pt]
+\frac{1}{4}\left[\left(\lambda+\frac{1}{\lambda}\right)\frac{{\cal Q}+I}{\sqrt{k+2}}+\left(\lambda+\frac{1}{\lambda}\right)\frac{\bar{I}}{\sqrt{k}}\right]^{2}\;.\end{array}\label{Ap4}\end{equation}
The middle lines in (\ref{Ap3}) and (\ref{Ap4}) express the effect
of the single constant magnetic field while the last lines - the gravitational
backreactions. ${\cal Q}_{i}$, $i=1,2,3$ refer to the helicity operators
corresponding to the internal left fermions in the background $SU(2)_{k}\times\mathbb{R}_{\phi}\times K^{6}$
and $j=0,1,2,...,p$ for $k=2p$.

Similarly, the right moving mass spectra read:

\begin{equation}
\begin{array}{c}
M_{R}^{2}=-1+\frac{P^{2}}{k_{g}}+\frac{(j+1/2)^{2}-({\cal Q}+I)^{2}}{k+2}+\bar{E}_{0}+\\[4pt]
+\frac{1+\sqrt{1+F^{2}}}{2}\left[\frac{{\cal Q}+I}{\sqrt{k+2}}+\frac{F\bar{{\cal P}}}{\sqrt{k_{g}}\left(1+\sqrt{1+F^{2}}\right)}\right]^{2}+\\[4pt]
+\frac{1}{4}\left[\left(\lambda+\frac{1}{\lambda}\right)\frac{{\cal Q}+I}{\sqrt{k+2}}+\left(\lambda+\frac{1}{\lambda}\right)\frac{\bar{I}}{\sqrt{k}}\right]^{2}\;.\end{array}\label{Ap5}\end{equation}
Together with the partition function $Z^{W}(\tau,\bar{\tau})$
for the curved 4-spacetime $W=SU(2)_{k=2p}\times\mathbb{R}_{\phi}$, as described in (\ref{Z-W}), we have the complete and exact string spectra in the presence of the constant
magnetic field and backreaction curvature, ${\cal R}$.

Now we can focus at a charged scalar particle moving through the 4d background such that quantum corrections due to magnetic field and its gravitational backreactions become valid. Exact field content of the string background $SU(2)_k\times \mathbb{R}_{\phi}$ in the case of magnetic and gravitational deformations (squashed 3-sphere), reads \cite{AsselmKrol2011f,KK95}:
\begin{equation}
\begin{array}{c}
G_{00}=1,\: G_{\beta\beta}=\frac{k}{4}\\[4pt]
G_{\alpha\alpha}=\frac{k}{4}\frac{(\lambda^{2}+1)^{2}-(8H^{2}\lambda^{2}+(\lambda^{2}-1)^{2})\cos^{2}\beta)}{(\lambda^{2}+1+(\lambda^{2}-1)\cos\beta)^{2}}\\[4pt]
G_{\gamma\gamma}=\frac{k}{4}\frac{(\lambda^{2}+1)^{2}-(8H^{2}\lambda^{2}-(\lambda^{2}-1)^{2})\cos^{2}\beta)}{(\lambda^{2}+1+(\lambda^{2}-1)\cos\beta)^{2}}\\[4pt]
G_{\alpha\gamma}=\frac{k}{4}\frac{4\lambda^{2}(1-2H^{2})\cos\beta+(\lambda^{4}-1)\sin^{2}\beta}{(\lambda^{2}+1+(\lambda^{2}-1)\cos\beta)^{2}}\\[4pt]
B_{\alpha\gamma}=\frac{k}{4}\frac{\lambda^{2}-1+(\lambda^{2}+1)\cos\beta}{(\lambda^{2}+1+(\lambda^{2}-1)\cos\beta)^{2}}\\[4pt]
A_{\alpha}=2g\sqrt{k}\frac{H\lambda\cos\beta}{(\lambda^{2}+1+(\lambda^{2}-1)\cos\beta)^{2}}\\[4pt]
A_{\gamma}=2g\sqrt{k}\frac{H\lambda}{(\lambda^{2}+1+(\lambda^{2}-1)\cos\beta)^{2}}\\[4pt]
\Phi=\frac{t}{\sqrt{k+2}}-\frac{1}{2}\log\left[\lambda+\frac{1}{\lambda}+(\lambda-\frac{1}{\lambda})\cos\beta\right]\end{array}\label{BCKG-G-H}\end{equation}
Again, the dependence on $\lambda$ shows the effect of gravitational backreaction of magnetic field, the $k$ dependance is due to the curvature of the 4d part of the background. 

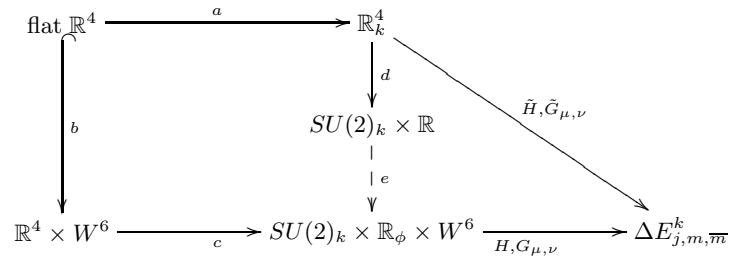
\begin{figure}[ht]
\centering
\begin{tabular}{c} 
\xymatrix@C=2pc@R=2pc{
\rm{flat}\; \mathbb{R}^4 \ar@{^(->}[dd]^{b} \ar[rr]^{a} & & \mathbb{R}^4_k \ar[d]^{d} \ar[ddrr]^{\tilde{H},\tilde{G}_{\mu,\nu}} \\ 
 & & SU(2)_k\times \mathbb{R} \ar@{-->}[d]^{e} \\ 
\mathbb{R}^4\times W^6 \ar[rr]_{c} & & SU(2)_k\times \mathbb{R}_{\phi}\times W^6 \ar[rr]_{H,G_{\mu,\nu}} & & \Delta E^k_{j,m,\overline{m}}}
\end{tabular}

\caption{$a$ is the change of smoothness on $\mathbb{R}^4$ from standard one to exotic $\mathbb{R}^4_k$; $b$ is the embedding of flat smooth $\mathbb{R}^4$ into the string background; $c$ is the change of the string backgrounds; $d$ assigns $\mathbb{R}^4_k$ $SU(2)_k\times \mathbb{R}$ the end of exotic $\mathbb{R}^4_k$, via GV invariant; $e$ is the embedding of $SU(2)_k\times \mathbb{R}$ into the string background; $H,G_{\mu,\nu}$ is the deformation of the CFT background resulting in the deformed spectrum $\Delta E^k_{j,m,\overline{m}}$; the same spectrum is obtained when $\tilde{H},\tilde{G}_{\mu,\nu}$ are on exotic $\mathbb{R}^4_k$} \label{fig-2}
\end{figure}

Such exact background allows for comparing it with 4d field theory calculations. When done, the scalar charged particle has modified energy spectrum, as follows \cite{AsselmKrol2011f,KK95}:
\begin{equation}
\Delta E_{j,m,\overline{m}}^{k}=\frac{1}{k+2}[j(j+1)-m^{2}]+\frac{(2\sqrt{k+2}eH-(\lambda+\frac{1}{\lambda})m-(\lambda-\frac{1}{\lambda})\sqrt{(1+2/k)}\overline{m})^{2}}{4(k+2)(1-2H^{2})}\:.\label{Mod-4d}\end{equation}
This result can be, however, interpreted directly as the spectrum of scalar, charged particle $e$, moving through exotic $\mathbb{R}^4_k$ geometry underlying the 4d spacetime. 
Such interpretation results from the following ingredients discussed in this paper:
\begin{itemize}
\item[i.] Let us start with flat, standard $\mathbb{R}^4$;
\item[ii.] include constant magnetic field, hence density of energy, into 4d flat space;
\item[iii.] background of closed superstring theory becomes curved and gravitational backreactions should be included;
\item[iv.] in heterotic (and type II) superstring theory the 4d flat part of the background is replaced by the curved according to: $\mathbb{R}^4\times W^6\to SU(2)_{k=2p}\times \mathbb{R}_{\phi}\times W^6$; 
\item[v.] the comparison of 4d field theory spectra with superstring theory in the deformed backgrounds, give rise to the deformed spectrum as in (\ref{Mod-4d});
\item[vi.] from the other side, starting with flat $\mathbb{R}^4$ we change its smooth structure to the exotic $\mathbb{R}^4_k$;
\item[vii.] the geometry of the end, $S^3\times \mathbb{R}$, and the connection of exotic $\mathbb{R}^4_k$ with codimension-1 foliations of $S^3$, gives rise to the WZ term of the $SU(2)_k$ WZW model;
\item[viii.] Exotic $\mathbb{R}^4_k$ is not flat hence contains a kind of gravity. This gravity in suitable limit should be quantized; the natural choice is to refer to WZW $SU(2)_k$ and string theory;
\item[ix.] supersymmetry, additional dimensions and string techniques should be considered as mathematics allowing for the consistent, from the point of view of QG, change between string backgrounds with different topologies.
\end{itemize} 
One can illustrate the above net of reasoning by the diagram in Fig. \ref{fig-2}. 
This is the first time when one is able to derive such definite calculations on small exotic $\mathbb{R}^4$. Moreover, such an approach shows that QG can be effectively formulated in 4d at least for the effects of gravity confined to exotic 4-geometry. String theory plays a role of  mathematics which was built especially for QG and the unification with other interactions. Hence, supersymmetry, additional dimensions etc. constitute ,,merely'' mathematical toolkit for exploring 4d QG and exotic smoothness on open manifolds. Still it would be extremely interesting to obtain QG results via path integral technique on exotic $\mathbb{R}^4_k$ and compare them with the above heuristic derivation via superstring theory. The work is in progress. 

\section*{Acknowledgment}
Based on the talk presented at Quantum Theory and Symmetries 7, Prague, August 7-13, 2011. I would like to thank Torsten Asselmeyer-Maluga for the numerous stimulating discussions and working together on the role of string theory in 4d exotic smoothness.



\end{document}